\documentclass[prl,amssymb,amsfonts,showpacks,twocolumn,showpacs]{revtex4}
\usepackage{amsmath}
\usepackage{amsfonts}
\usepackage{amssymb}
\usepackage{bm}

\def\beq{\begin{equation}}
\def\eeq{\end{equation}}
\def\bea{\begin{eqnarray}}
\def\eea{\end{eqnarray}}

\def\pad{\partial}
\def\tr{{\rm tr}}

\def\cF{{\cal F}}
\def\cH{{\cal H}}

\def\etal{{\it et al.\/}}

\def\1{\mbox{1\hskip-.25em l}}
\def\6{\langle }
\def\9{\rangle}
\usepackage{graphicx}
\usepackage{epsfig}
\begin{document}
\title{Entropy, holography and the second law}
\author{Daniel R. Terno}
\affiliation{Perimeter Institute for Theoretical Physics, 35 King St. N., Waterloo,
Ontario, Canada N2J 2W9}
\begin{abstract}
The geometric entropy in quantum field theory is not a Lorentz
scalar and has no invariant meaning, while the black hole entropy
is invariant. Renormalization of entropy and energy for reduced
density matrices may lead to the negative free energy even if no
boundary conditions are imposed. Presence of particles outside the
horizon of a uniformly accelerated observer prevents the
description in terms of a single Unruh temperature.
\end{abstract}
\pacs{04.70.Dy, 03.67.-a, 04.60.-m}
\maketitle
Entropy plays a central role in statistical physics,
thermodynamics and information theory
\cite{ll5,wehrl,coto,ap:b,pt:04}. The von Neumann entropy of a quantum
system that is described by a density operator $\rho$,
\beq
S=-\tr \rho \log \rho \label{vN},
\eeq
has a simple information-theoretical meaning. Each  measurement
scheme leads to a probability distribution $\{p_\mu\}$ on its
outcomes, and its Shannon's entropy is
\beq
H=-\sum_\mu p_\mu\log p_\mu.
\eeq
The von Neumann entropy  of the state $\rho$ is the lowest value
that can be reached by the Shannon entropy over all possible
probability distributions. The concepts of quantum information
theory, with suitable adjustments, can be applied to any quantum
system, including field theories on a curved background
\cite{pt:04,t:02}.

To connect von Neumann's entropy with that of thermodynamics, a
number of assumptions should be made \cite{ll5,ap:b,wehrl,el:b}.
They are natural for a normal matter, and there are no
experimental results that may hint at their possible violation.
However, rigorous proofs exist only for a small number of simple
models
\cite{wehrl,el:b}.

Black holes are  usually considered thermodynamical systems in
their own right \cite{bh,fn:b}. Their  entropy  is one quarter of
their horizon area in Plank units,
\beq
S_{\rm BH}=A^2/l_P^2,\label{bhs}
\eeq
and their temperature is proportional to the surface gravity
\cite{bek,ha:75}. This connection between area and entropy is elevated
to a general physical law in different versions of the holographic
principle \cite{fn:b,busso:rmp} that bounds the  entropy inside a
volume by the quarter of its area (the definitions of entropy and
area vary, however). Perhaps the most striking consequence of the
proportionality of entropy and  area is the
 derivation \cite{jac:95} of the Einstein equations  from the
holographic principle together with the thermodynamic relation
$\delta Q=TdS$.

The notion of  reduced density operator and its entropy is
fundamental for  entanglement theory and quantum information in
general
\cite{ap:b,pt:04}. It is obtained by tracing out the irrelevant degrees
of freedom (in the Hilbert space language) \cite{ap:b} or by
restricting  states to the local algebra of  observables (in the
framework of algebraic quantum theory) \cite{haag}. For pure
states the von Neumann entropy of a reduced density operator is
 the measure of  entanglement between the degrees of freedom ``here" and
  ``elsewhere". This is  what is meant by the geometric
  entropy \cite{gentr,notes1}. The actual calculations of  reduced density
  operators are quite complicated. Only the analysis for a half-space in the
  Minkowski spacetime (where $\rho$ can be obtained from the Rindler
quantization and is related to the Unruh effect \cite{bh,unruh})
is relatively straightforward.
  The geometric entropy usually diverges. However, it was shown
  in a number of cases \cite{ gentr, bom:86,yur:03}
  that   the regularized geometric entropy is
  proportional to the boundary area.

In this Letter we discuss some properties of entropy and its
relation to thermodynamics. We show that the geometric entropy is
not a Lorentz scalar for essentially the same reasons as the spin
entropy of massive particles \cite{pt:04}. We consider the
implications of the presence of Minkowski particles beyond the
horizon of an accelerated observer.  Her temperature is defined
\cite{ll5} by
\beq
\frac{1}{T}=\frac{\pad S}{\pad E}, \label{temper}
\eeq
using the renormalized energy and entropy. 
This temperature is different from the Unruh temperature for the
same proper acceleration. We furthermore connect these results to
the generalized second law, the derivation of the Einstein
equations from thermodynamics and the microscopic origins of the
black hole entropy.

Let us assume that  splitting into ``inside" and ``outside"
degrees of freedom allows writing a Fock space of some free
quantum field theory as
$\cF(\cH)\simeq\cF_1(\cH_1)\otimes\cF_2(\cH_2)$. This
decomposition may be associated either with spatial regions on a
given time slice or with spacetime domains. The Lorentz-invariant
vacuum state
\beq
|\Omega\9=\sum_{ij}c_{ij}|\psi_i^1\9\otimes|\psi_j^2\9,
\eeq
has an explicit entangled form \cite{pt:04,bom:86,takagi}, and
this is also true  for one- and many-particle states. The reduced
density matrix that corresponds to  tracing out the ``outside"
degrees of freedom (that are labelled by ``2") is
$\rho^1_{ij}=\sum_k c_{ik}c^*_{jk}$.

{\em Proposition.} The reduced density matrix $\rho_{ij}$ is not
covariant under Lorentz transformations.

For $\rho$ to have  a definite transformation law, it  is
necessary for a unitary transformation on the entire space,
$U(\Lambda)$, to be a direct product of unitaries that act on the
``inside" and ``outside" spaces, $U(\Lambda)=U_1(\Lambda)\otimes
U_2(\Lambda)$. The vacuum state spans a one-dimensional
irreducible representation space of the Poincare group. The direct
product structure is compatible with this property of $|\Omega\9$
only if both representations $U_1$ and $U_2$ are one-dimensional.
However, the ensuing product structure of the vacuum  is
incompatible with the long-range correlations in it, and, in
particular, with the violation of the Bell-type inequalities
\cite{pt:04,corr}. A similar argument applies to one-particle
states, which are faithful irreducible representations of the
Poincare group and, therefore, cannot be described as a direct
product of the two representations.

Hence, unlike the full state that transforms unitarily, its
reduced density matrices do not have a definite Lorentz
transformation law, and the entropy is  not necessarily invariant.
A byproduct of the non-invariance of entropy is that the
(effective) number of degrees of freedom, $N=e^{S_{\max}}$, is
frame-dependent.

Consider a massless scalar field in a box of the size $L$ in
Minkowski spacetime. Compliance with the spacelike holographic
bound \cite{yur:03}
   is enforced by imposing  two
 cut-offs. First, one must limit the maximal frequency of each mode
at about the Planck frequency $1/l_P$. Then, in order to ensure
  gravitational stability  only the states below a
certain maximal energy are allowed.  This bound is expressed (in
the suitable units) as a dimensionless ratio $B\sim L/l_P$. As a
result, in the rest frame of the box the maximal entropy is finite
and proportional to the surface area, $S_{\max}=\ln N\propto
L^2/l_P^2$,
\cite{yur:03}.

Now  consider another observer that moves along one of the box's
edges with some velocity $v>0$. In his frame the surface area is
$A'=2L^2(1+2/\gamma)$, where $\gamma=1/\sqrt{1-v^2}$. The number
of admissible modes should be lowered by about a factor of
$\gamma(1+v)$ because of the Doppler effect. Hence, following the
derivations of Ref.~\cite{yur:03} the new maximal entropy is
\beq
S'_{\max}=S_{\max}/\sqrt{\gamma(1+v)}.
\eeq
 In this example the maximal entropy is reduced faster than
the surface area, so the spacelike holographic bound is satisfied
for all velocities $v$.
\begin{figure}[htbp]
\epsfxsize=0.43\textwidth
\centerline{\epsffile{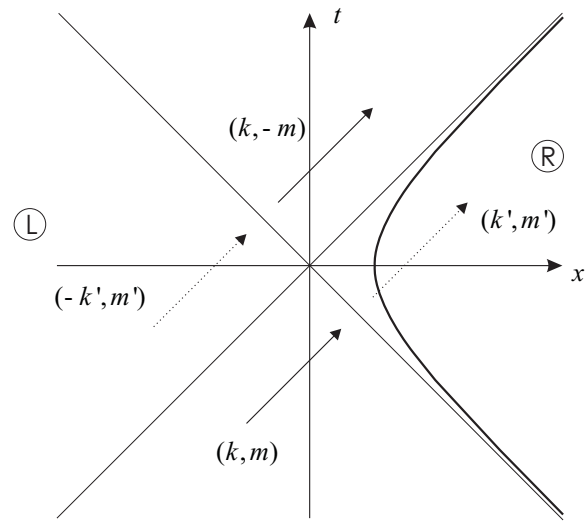}} \vspace*{-0.2cm} \caption{\small{
Hyperbolic trajectory of a uniformly accelerated observer is shown
in a thick line. With the standard choice of reference frames
Alice crosses the $x$-axis at $x=1/a$.
 Her past and future horizons define the right and  left Rindler
wedges. Solid arrows represent Minkowski particles with the
trajectory parameters $(k,\pm m)$, while the dotted arrows
represent the same particles as described by Alice. The mode in
our example corresponds to $(k,-m)$ arrow.}}
\vspace*{-0.4cm}
\end{figure}

An isolated stationary black hole furnishes an example of the {\em
invariant} entropy.  Consider two observers  at the same spacetime
point outside the horizon, but not necessarily in the
asymptotically flat region. Their local coordinate frames are
connected by some Lorentz transformation $\Lambda$. The global
role of this local Lorentz transformation is
 to define  new
surfaces of simultaneity with respect to the Lorentz-transformed
observer. These surfaces may intersect the event horizon
differently from those of the original observer. However, the area
theorem of Hawking
\cite{fn:b} guaranties that all these intersections lead to the same
horizon area.  The semiclassical  black hole entropy is given by
Eq.~(\ref{bhs}) and the leading corrections are  functions of the
area only
\cite{gour}. Hence the invariance
of the horizon area keeps the entropy invariant. Since we are
dealing here with a curved spacetime, there is no contradiction
with the proposition that has
been proven above. %

 The relations between entropy, energy
and temperature are important in at least two contexts. The
derivation
\cite{jac:95} of the Einstein equations from the holographic
principle and thermodynamics requires an independent expression
for the temperature. It is taken to be the Unruh temperature of
 an accelerated observer in vacuum, while the matter is
located beyond the observer's horizon. It is  argued
\cite{notes1,maso:03} that a necessary condition for the
generalized second law
\cite{bh,fn:b}   is  that there are no highly
 entropic  objects, i.e.,
\beq
S\leq E/T, \label{cond}
\eeq
should hold, where $T$ is the black hole temperature and $E$ and
$S$ are object's energy and entropy. This is to say that object's
free energy  is non-negative, $F=E-TS\geq 0.$

Consider a free real massless scalar field in $1+1$ Minkowski
spacetime from the point of view of a uniformly accelerated
observer (whom we conventionally name Alice). Minkowski particles
are present beyond Alice's horizon in the left Rindler wedge.
 We investigate the relationship between the renormalized
entropy, the renormalized energy and the temperature as observed
by Alice (see Fig.~1 for definitions).  Thanks to the results of
Audretsch  and M\"{u}ller
\cite{am:94}  all the relevant
quantities can be calculated explicitly.  Localized particles are
most conveniently described in the wave packet basis
\cite{ha:75,takagi}.
The Minkowski wave packet is defined by the superposition of the
positive energy plane wave solutions
$f_k=\frac{1}{\sqrt{4\pi\omega_k}}e^{-i(\omega_k t-kx)}$, where
$\omega_k=|k|$, in a momentum interval $\epsilon$,
\beq
f_{km}(t,x)=\frac{1}{\sqrt{\epsilon}}
\int_k^{k+\epsilon}d\tilde{k} e^{-i\tilde{k}m}f_{\tilde{k}}(t,x).
\eeq
For each wave packet both $k/\epsilon$ and $m\epsilon/2\pi$ are
fixed integers. These wave packets form a complete orthonormal
basis,  and the quantum field is written with $a_{km}$ and
$a_{km}^\dag$ operators.  For a small value of $\epsilon$, the
state $|k,m\9=a_{km}^\dag|\Omega\9$ represents a particle with
energy $\6E\9\approx\omega_k$ whose spatial localization  roughly
corresponds to the position of the maximum of the wave packet
$f_{km}$. All the reasonable localization POVM's \cite{pt:04} give
only a power-law decrease of the probability density to find a
massless particle at some distance from the classical trajectory
that is specified by $m$. In particular, it is impossible to
confine any Minkowski state exactly to one of the Rindler wedges.

The right and left Rindler modes $f^R_{k'm'}, f^L_{k'm'}$ are
defined similarly. Fig.~1 represents a correspondence between the
energy-momentum vector of a Minkowski particle,
$k^\mu=(\omega_k,k)$ in Minkowski and Rindler reference frames.
For $k>0,m>0$ the equivalent description is given by
\beq
(k',m')\Leftrightarrow
(k,m)=(e^{-am'}k',e^{am'}/a),\label{equmode}
\eeq
where $a$ is Alice's proper acceleration. The Bogoliubov
coefficients are non-zero essentially only between the equivalent
modes.

The Minkowski vacuum then takes the form
\beq
|\Omega\9=\prod_{
\tilde{k}', \tilde{m}'}\frac{1}{|\alpha_{\tilde{k}'}|}
\sum_{q=0}^{\infty}\left(\frac{\beta_{\tilde{k}'}^*}{\alpha_{\tilde{k}'}}\right)^q
|q_{-\tilde{k}'\tilde{m}'}\9_L\otimes
|q_{\tilde{k}'\tilde{m}'}\9_R,
\eeq
where $|q_{-\tilde{k}'\tilde{m}'}\9_L$ is a state with $q$
particles in the mode $-\tilde{k}'\tilde{m}'$ in the left Rindler
wedge, etc. The coefficients $\alpha_{\tilde{k}'}$,
$\beta_{\tilde{k}'}$ are given in \cite{am:94}. Accordingly, the
reduced density matrices in both wedges are
\beq
\rho_\Omega^{L,R}=\prod_{
\tilde{k}',
\tilde{m}'}\frac{1}{Z_{\tilde{k}'}}\sum_{q=0}^{\infty}e^{-\omega_{\tilde{k}'}\beta q}
|q_{\mp\tilde{k}'\tilde{m}'}\9\6 q_{\mp\tilde{k}'\tilde{m}'}|,
\eeq
where $Z_{\tilde{k}'}=(1-\exp(-\omega_{\tilde{k}'}\beta))^{-1}$,
and the parameter $\beta$ defined by the  proper acceleration,
$\beta=2\pi/a. $ This is a thermal distribution with
$T_U=1/\beta$.

  A state
$|n_{km}\9$ with $n$ Minkowski particles that (essentially) pass
through the {\em left} wedge has a more complicated form
\cite{am:94}. After some approximations, its reduced density matrix in the right wedge is
\begin{widetext}
\beq
\rho_{(n;km)}^R=\prod_{\tilde{k}'\neq k', \tilde{m}'\neq m'}
\frac{1}{Z_{\tilde{k}'}}\sum_{q=0}^{\infty}e^{-\omega_{\tilde{k}'}\beta q}
|q_{\tilde{k}'\tilde{m}'}\9\6
q_{\tilde{k}'\tilde{m}'}|\otimes\frac{1}{Z_{k'}^{n+1}}\sum_{r=0}^\infty\frac{(n+r)!}{n!r!}
e^{-\omega_{k'}\beta r}|r_{k'm'}\9\6 r_{k'm'}|, \label{rhor}
\eeq
\end{widetext}
where the ``mirror mode" $(k',m')$ is related to $(k,-m)$ by Eq.~
(\ref{equmode}). The number of excited modes is infinite, so the
entropy diverges. A finite renormalized entropy of $\rho^R$ is
defined as follows \cite{gentr,notes1}. First the regularized
entropies of $\rho_\Omega^R$ and $\rho^R$ are calculated by
imposing a cutoff $l$. Then the difference
$S(\rho^R;l)-S(\rho_\Omega;l)$ of the cutoff-dependent entropies
is taken. Finally the  cutoff is removed. The renormalized energy
is calculated similarly, with the Hamiltonian taken with respect
to the Rindler time
\cite{unruh,notes1}. Both quantities are independent of the
cut-off, but their non-zero values
 result from the modes influenced
by Minkowski particles.

 In the limit $\exp(\beta\omega)\gg n$ it is easy to get explicit
 expressions for $S$ and $E$.
 For simplicity we consider the
 case with only one  exited Minkowski mode.
In this limit
\beq
S=(\omega\beta n +n-(n+1)\ln(n+1))e^{-\beta\omega}, \label{s1}
\eeq
and
\beq
E=\omega n e^{-\beta\omega}, \label{e1}
\eeq
where the energy $\omega_{k'}$ of the equivalent Rindler mode
$(k',m')$ is denoted  by $\omega$.

The notion of temperature is unapplicable for  systems far from
equilibrium, so the thermodynamic description is valid only when
$\rho^R$ is well approximated by $\exp(-H/T)/\tr \exp(-H/T)$. In
the free field theories the modes are independent. There is no way
in which the Unruh temperature of the Minkowski vacuum,
$T_U=1/\beta=a/2\pi$, is relevant for the modes like $(k',m')$.
Hence the correct interpretation  is to split the system into two
 non-interacting parts. All the modes except for
$(k',m')$  are thermal with a usual Unruh temperature $T_U$ form
one subsystem. The mode $(k',m')$ forms another. When $n
\exp(-\beta \omega)\ll 1$, this mode behaves as a two-level system and its
density matrix is close to the thermal distribution with the
temperature $T$ that is introduced according to 
the definition in Eq.~(\ref{temper}). A variable parameter is $n$,
a number of the Minkowski particles in a given mode. While
fractional expected number of particles between $n$ and $n+1$ is
accomplished by mixing states with these particle numbers, it is
consistent  to vary $n$ as a real parameter in the approximation
of Eqs.~(\ref{s1}). The temperature is
\beq
\frac{1}{T}=\frac{dS/dn}{dE/dn}=\beta-\frac{\ln{(n+1)}}{\omega}. \label{t1}
\eeq
 As expected, with no Minkowski particles present, the temperature
reduces to the Unruh temperature $T_U=1/\beta$.  It is easy to see
that for the state $\rho_{(n;km)}^R$
\beq
S-E/T=(n-\ln(n+1))e^{-\beta\omega}>0,
\eeq
i.e., its free energy is negative. It is consistent with
$F=-E/3<0$  for the ``usual" black body radiation
\cite{ll5}.

Like in the field-in-a-box example, the  entropy (\ref {s1}) is
not invariant under Lorentz transformations. Consider now another
accelerated observer (Bob). At the Minkowski time $t=0$ Alice
passes through $x=1/a$ with zero velocity (see Fig.~1). Bob has
the same constant proper acceleration and  passes thought the same
spacetime point, but with the velocity $v>0$. The Rindler wedges
of Bob and Alice overlap, but are not identical. In construction
of their reduced density matrices different parts of the Universe
are excluded, so their descriptions are not compatible
\cite{pt:02}.
 The easiest way to
get Bob's description of the quantum state $|n_{km}\9$ is first to
pass to the Minkowski frame where Bob's trajectory is that of the
standard accelerated observer (i.e., looks like Alice's trajectory
on Fig.~1). Then an application of  Bogoliubov transformations
gives Bob's state. Since Minkowski vacuum is Lorentz-invariant, it
looks to Alice the same way as to Bob. However, the particles look
differently: in the new Minkowski frame their energy acquires a
Doppler factor $\sqrt{(1-v)/(1+v)}$, and the trajectory parameter
$m$ is increased to reflect the change in the frame's origin.
Using Eq.~(\ref{equmode}) it is easy to show that depending on the
relative magnitude of $am$ and $v$ the Rindler energy of the
relevant Bob's mode is either greater or smaller than that of the
corresponding Alice's mode.
 A substitution to Eqs.~(\ref{s1}) and
(\ref{t1}) shows that $S_B\neq S_A$, $T_B\neq T_A$ as well.

We showed that in Minkowski spacetime entropy of a reduced density
matrix is not necessarily invariant quantity, so the entanglement
between different degrees of freedom and their effective numbers
are observer-dependent. Taking this non-invariance into account
weakens the arguments against a spacelike entropy bound
\cite{busso:rmp}, since they  are based on  the
invariance of entropy. Presence of the matter fields outside the
horizon may either change Unruh temperature or  make thermodynamic
description altogether inapplicable. Corrections to the Unruh
temperature may point to the quantum corrections to the Einstein
equations. On the other hand, the black hole entropy is invariant.
This invariance may help to pick up the correct explanation of the
black hole entropy from a multitude of different approaches
\cite{pt:04,fn:b}.

This work was initiated by discussions with Florian Girelli and
David Poulin. Helpful suggestions and critical comments of Jacob
Bekenstein, Netanel Lindner, Rob Myers, Amos Ori, Terry Rudolph,
Lee Smolin,
 Rafael Sorkin and Rowan Thompson are gratefully acknowledged.

\end{document}